\documentclass[prb,aps,twocolumn]{revtex4}
\usepackage{graphicx,amssymb,amsmath,color,psfrag}
\usepackage{amsthm}
\usepackage{amsfonts}
\usepackage{algorithmic}
\usepackage{enumerate}
\usepackage{latexsym}

\begin{document}
\newcommand{\dt}{\Delta\tau}
\newcommand{\al}{\alpha}
\newcommand{\ep}{\varepsilon}
\newcommand{\ave}[1]{\langle #1\rangle}
\newcommand{\have}[1]{\langle #1\rangle_{\{s\}}}
\newcommand{\bave}[1]{\big\langle #1\big\rangle}
\newcommand{\Bave}[1]{\Big\langle #1\Big\rangle}
\newcommand{\dave}[1]{\langle\langle #1\rangle\rangle}
\newcommand{\bigdave}[1]{\big\langle\big\langle #1\big\rangle\big\rangle}
\newcommand{\Bigdave}[1]{\Big\langle\Big\langle #1\Big\rangle\Big\rangle}
\newcommand{\braket}[2]{\langle #1|#2\rangle}
\newcommand{\up}{\uparrow}
\newcommand{\dn}{\downarrow}
\newcommand{\bb}{\mathsf{B}}
\newcommand{\ctr}{{\text{\Large${\mathcal T}r$}}}
\newcommand{\sctr}{{\mathcal{T}}\!r \,}
\newcommand{\btr}{\underset{\{s\}}{\text{\Large\rm Tr}}}
\newcommand{\lvec}[1]{\mathbf{#1}}
\newcommand{\gt}{\tilde{g}}
\newcommand{\ggt}{\tilde{G}}
\newcommand{\jpsj}{J.\ Phys.\ Soc.\ Japan\ }

\title{Magnetic Impurity in Bernal Stacked Bilayer Graphene}
\author{J. H. Sun$,^{1}$ F. M. Hu$,^2$ H. K. Tang$,^{1}$ and H. Q. Lin$^1$}
\affiliation{$^{1}$ Department of Physics and ITP, The Chinese University of Hong Kong, Hong Kong, China\\
$^2$COMP/Department of Applied Physics, Aalto University School of Science, P.O. Box 11000, FI-00076 Aalto, Espoo, Finland}

\begin{abstract}
We investigate a magnetic impurity in Bernal stacked bilayer graphene by a non-perturbative numerical exact approach. In the two cases we study, impurity is placed on the top of two different sublattices (A and B) in bilayer graphene. We find that similar to the monolayer case, magnet moment of the impurity could still be tuned in a wide range through changing the chemical potential. However, the property of the impurity depends strongly on its location due to the broken symmetry between sublattices A and B caused by the Bernal stacking. This difference becomes more apparent with the increase in the hybridization and decrease in the on-site Coulomb repulsion. Additionally, we calculate the impurity spectral densities and the correlation functions between the impurity and the conduction-band electrons. All the computational results show the same spatial dependence on the location of the impurity.
\end{abstract}

\date{\today}
\maketitle

\section{Introduction}

Following the fabrication of monolayer graphene (MLG), \cite{Novoselov04} bilayer graphene(BLG) has attracted intensive attention due to its unique electrical flexibility \cite{Castro07,Oostinga07,Zhang09,Mak09,Jr12} and unusual physical properties including the unconventional quantum Hall effects. \cite{Zhang05,Novoselov06,McCann06,Zhangfan12} At the same time, BLG is regarded as a good candidate in spintronics because compared with MLG, BLG shows longer spin-relaxation times at room temperature, \cite{Han11, Yang11} which is crucial for spintronic devices. \cite{Zutic04} In particular, transition metal atom is usually used as a spin provider in spintronic devices, \cite{Pesin12} and this motivates us to study the properties of magnetic adatom in the BLG system.

Because of the vanishing density of states (DOS) at the Dirac point, when we dope MLG with magnetic adatoms, their local moments could be conserved well at finite temperature. \cite{Sengupta08} The behavior of magnetic impurity in MLG has been considerably studied \cite{Uchoa08,Nieminen09,Cornaglia09,Hu11} and Kondo effects are difficult to observe in this system. Significantly different from the MLG, the two sublattices in Bernal stacked\cite{Rabii82} BLG are no longer equivalent. The DOS near the Dirac point would be largely modified \cite{Castro09} because of the degeneracy lifted by the inter-layer hoppings and the local density of states (LDOS) have distinguishing values on the two sublattices which plays a more important role. This spatial inhomogeneity in LDOS could be directly detected experimentally with the scanning tunneling microscopy (STM). \cite{Ugeda10} In BLG with Bernal stacking, it has been reported that whether the missing carbon atoms (vacancy) are generated on sublattice A or B greatly alters the defect-induced magnetism and the localization of the zero mode in vacancy states. \cite{Ugeda10,Castro10} In addition, the use of STM tip provides the possibility to control the position of adatoms with atomic precision on the two-dimensional open surface. \cite{Eigler90,Brar11}

\begin{figure}[t]
\begin{center}
\includegraphics[scale=0.45, bb=16 65 605 309]{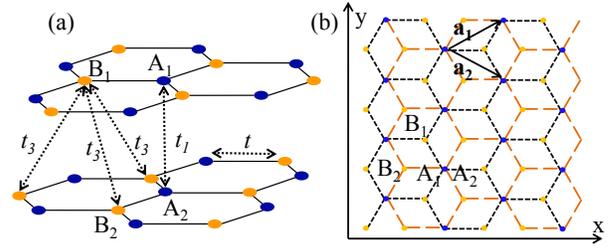}
\end{center}
\caption{(Color online). (a) Lattice structure and various hopping energies of Bernal stacked BLG . (b) Top view of BLG with the black(dotted) line forming the top layer and the orange(dashed) line forming the bottom layer. $\bf{a_1}$ and $\bf{a_2}$ are the surface unit vectors.  }\label{Fig:blg}
\end{figure}

In Fig.~\ref{Fig:blg}, we show such asymmetry: the carbon atoms on sublattice A belonging to layer $1$ lies directly on the top of those belonging to layer $2$ so that the number of inter-layer hopping is one, while the carbon atoms on sublattice B in one layer is just fixed at the center of hexagon in the other layer so that it has three inter-layer hoppings. The direct consequence is that the LDOS in Bernal stacked BLG has spatial inhomogeneity. In the vicinity of Dirac point, the LDOS on sublattice A is much smaller than that on sublattice B. We find that the local moment of the magnetic impurity on top of sublattice A could develop better than that on sublattice B. This difference is not only reflected by the physical quantities of impurity itself but also by the spatial correlation functions between the impurity and the carbon electrons.

In this paper, we report a quantum Monte Carlo (QMC) study of a magnetic impurity placed on the top of the sublattices A and B in the Bernal stacked BLG and compare the results with their monolayer counterpart. The QMC method we used deals with infinite sea of conduction electrons and many-body effect without any approximations, so the results we present are essentially exact. The paper is organized as follows. In Sec. II, we give a brief introduction to our theoretical model, the Anderson impurity model in Bernal stacked BLG, as well as our numerical methods. We also compute LDOS for sublattices A and B in BLG without impurity and compare them with the monolayer case. The purpose is to see the symmetry breaking generated by the Bernal stacking in bilayer system. In Sec. III, we present our results obtained from the quantum Monte Carlo (QMC) simulation. Firstly, we show the basic thermodynamic quantities on the impurity site, including occupancy, double occupancy, local magnet moments and spin susceptibility varying the chemical potential and the temperature. Secondly, using the maximum entropy method with particle-hole symmetry, we do analytical continuation for the Green's function obtained from QMC to extract impurity's spectral densities. Finally, we study the charge-charge and spin-spin correlation functions between the impurity and the carbon sites. All the calculations were worked out on both sublattices A and B and the results in bilayer case are compared with those in MLG. In Sec. IV, we summarize our results.

\section{Model and Methods}
 Our starting point is the single-impurity Anderson model which has an impurity orbit of energy $\varepsilon_{d}$ and on-site Coulomb repulsion $U$. \cite{Anderson61} The impurity orbit hybridizes with a conduction band with strength $V$. The total Hamiltonian is
\[
H=H_{0}+H_{1}+H_{2}.
\]
$H_0$ is a tight-binding Hamiltonian. For BLG it is

\begin{equation}
\begin{aligned}
H_0 &= -t\sum\limits_{<i,j>,m,\sigma}a_{mi\sigma}^\dagger b_{mj\sigma}+ \text{H.c.} \\
    & -t_1 \sum\limits_{j,\sigma} a_{1j\sigma}^\dagger a_{2j\sigma}+ \text{H.c.}\\
    &-t_3 \sum\limits_{<i,j>} b_{1i\sigma}^\dagger b_{2j\sigma} + \text{H.c.}\\
    &-\mu \sum\limits_{i,m,\sigma}( a_{mi\sigma}^\dagger a_{mi\sigma}+b_{mi\sigma}^\dagger b_{mi\sigma}) ,\\
\end{aligned}
\end{equation}
where $a_{m i\sigma}$($b_{m i\sigma}$) annihilates an electron with spin $\sigma$ at the site $R_{mia}$ ($R_{mib}$) on sublattice A(B) of graphene's hexagonal structure in the m-th layer. The lattice structure of BLG is shown in Fig.~\ref{Fig:blg}. Intra-layer hopping $t$ is the nearest neighbor hopping integral between sublattices A and B in the same plane and $t$ is about $2.8eV$. \cite{Rmp}  Here $t$ is used as the energy unit in our calculation. For the inter-layer hopping, we use Slonczewski-Weiss-McClure parametrization, \cite{Brandt88,Dresselhaus02} i.e., $t_1\approx0.4eV$ and $t_3\approx0.3eV$ and the configurations of these two couplings can be seen in Fig.~\ref{Fig:blg}(a).
$\mu$ is chemical potential and is equal to zero in pure graphene. \\
$H_1$ is the impurity Hamiltonian
\begin{equation}
\begin{aligned}
H_{1}=\sum_{\sigma}(\varepsilon_{d}-\mu)d^{\dag}_{\sigma}d_{\sigma}^{}+Ud^{\dag}_{\uparrow}d_{\uparrow}^{}d^{\dag}_{\downarrow}d_{\downarrow}^{},
\end{aligned}
\end{equation}
where $d_{\sigma}$ annihilates an electron with spin $\sigma$ at the impurity orbit.\\
Finally, $H_{2}$ describes the hybridization between the impurity adatom and a carbon atom on the BLG.
If we place the impurity adatom on the top of a site in the first layer, then the Hamiltonian is written as
\begin{equation}
\begin{aligned}
H_{2}= \sum_{\sigma}V(c_{10\sigma}d^{\dagger}_{\sigma}+\text{H.c.})\texttt{.}
\end{aligned}
\end{equation}

\begin{figure}[t]
\begin{center}
\includegraphics[scale=0.45, bb=90 45 465 410]{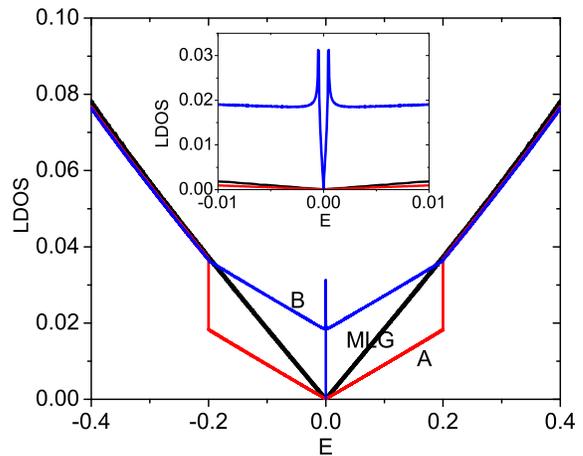}
\end{center}
\caption{(Color online). LDOS in the vicinity of Dirac point, from top to down is for sublattice B in BLG, MLG and sublattice A in BLG, respectively. The inset shows the detail of LDOS in the vicinity of Dirac energy. We set $t_1=0.2t$, $t_3=0.1t$ in BLG. } \label{Fig:dos}
\end{figure}

If the adatom is added on the top of sublattice A, $c_{10\sigma}=a_{10\sigma}$, otherwise $c_{10\sigma}=b_{10\sigma}$.

Our principal computational tool is the Hirsch-Fye QMC algorithm. \cite{hirsch86} Basic thermodynamic properties of the impurity site are directly calculated by this algorithm and the spectral density is extracted by the method of Bayesian statistical inference.
The Hirsch-Fye algorithm naturally returns the imaginary-time Green's function $G_d(\tau)= \sum_\sigma  {G_{d\sigma}  \left( \tau  \right)} $ of the impurity. With this Green's function, we can easily compute basic thermodynamic quantities of the impurity orbit, such as the expectation values of the total charge
\begin{equation}
\begin{aligned}
n_d=\langle n_{d\uparrow}+n_{d\downarrow}\rangle ,
\end{aligned}
\end{equation}
the local moment squared
\begin{equation}
\begin{aligned}
m_d^2=\langle (n_{d\uparrow}-n_{d\downarrow})^2 \rangle ,
\end{aligned}
\end{equation}
 and the double occupancy
 \begin{equation}
\begin{aligned}
n_{d\uparrow}n_{d\downarrow}=\langle n_{d\uparrow}n_{d\downarrow} \rangle .
 \end{aligned}
\end{equation}
According to the fact that the impurity charge can either be zero or one, we note that
\begin{equation}\label{eq:moment}
m_d^2=n_d-2n_{d\uparrow}n_{d\downarrow} .
\end{equation}
A non-zero value of $m_d^2$ indicates the formation of a moment on the adatom orbit. The closer this value is to one, the more fully developed is the moment. We also calculate the static impurity spin susceptibility
\begin{equation}\label{eq:suscep}
\chi=\int^{\beta}_{0}d\tau\langle m_{d}(\tau)m_{d}(0)\rangle,
\end{equation}
where $\beta=T^{-1}$, $m_{d}(\tau)=e^{\tau H}m_{d}(0)e^{-\tau H}$.

Using imaginary-time Green's function obtained from the QMC method, we can calculate the spectral density $A(\omega)=\sum_\sigma A_\sigma(\omega)$ by numerically solving \cite{jarrell96}
\begin{equation}
\begin{aligned}
G_d\left( \tau  \right)  ={\int\limits_{ - \infty }^\infty  {d\omega } } \frac{e^{  -\tau\omega}{A  \left(\omega  \right)}}
{{e^{ - \beta \omega}  + 1}}.
\end{aligned}
\end{equation}
The detailed procedure of Bayesian inference method is presented in Ref.~\onlinecite{jarrell96}. This Bayesian inference procedure is also called the maximum entropy method.

We also use an extension of the Hirsch-Fye algorithm \cite{Gubernatis87} to compute the charge-charge correlation function
\begin{equation}
\begin{aligned}
C_i = \langle n_d n_i\rangle -\langle n_d\rangle\langle n_i\rangle ,
\end{aligned}
\end{equation}
and the spin-spin correlation function
\begin{equation}
\begin{aligned}
S_i = \langle m_d m_i\rangle .
\end{aligned}
\end{equation}

Using the standard particle-hole transformation on one of the sublattice, we can prove that $G_d(\tau,\mu)=G_d(-\tau,-\mu)$ and consequently that $A_{\sigma}(\mu,\omega)=A_{\sigma}(-\mu,-\omega)$  when $\varepsilon_d=-U/2$. \cite{hu2012} These two results in turn imply the symmetries with respect to the sign of $\mu$ in various thermodynamic quantities of interest. These symmetries also mean that without loss of generality, we can restrict our attention to the behavior of the system when $\mu \le 0$.

The DOS of MLG and BLG are distinguishing in the vicinity of the Dirac point because of their different dispersion relations. \cite{Nilsson08}. In general, the degeneracy near the Dirac energy are lifted due to the inter-layer hoppings $t_1$ and $t_3$, so BLG shows larger DOS than MLG near the Dirac energy. \cite{Castro09}
To gain some primary insights of the spatial inhomogeneity in BLG, we first calculate the LDOS without the magnetic impurity ($V=0$). The results are shown in Fig.~\ref{Fig:dos} as well as the LDOS for monolayer case. In MLG, every site is identical so that the DOS is equal to the local one while in BLG the asymmetry between two sublattices occurs. In particular, as is shown in Fig.~\ref{Fig:dos}, it is clear that the LDOS on sublattice B has a larger value than that on sublattice A and the value of monolayer is in between. As a result, we can expect that the difference in LDOS for two sublattices will lead to position-dependent effective hybridizations between impurity and carbon atoms, and the features of impurity have spatial inhomogeneity which can be seen from QMC results in the following sections.

\section{Results}
\subsection{Basic Thermodynamic Quantities}

\begin{figure}[t]
\begin{center}
\includegraphics[scale=0.50, bb=56 65 565 380]{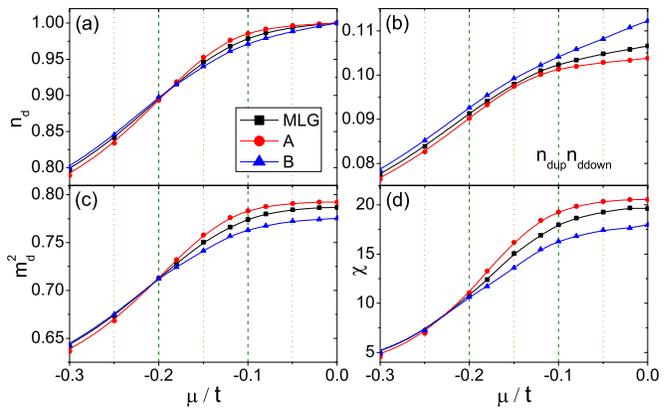}
\end{center}
\caption{(Color online). (a) $n_{d}$ as a function of $\mu$. (b) $n_{dup}n_{ddown}$ as a function of $\mu$. (c) $m_{d}^{2}$ as a function of $\mu$. (d) $\chi$ as a function of $\mu$. In all plates, $U=1.6t$, $\varepsilon_{d}=-U/2$, $\beta=1/T=40t^{-1}$, $V=1.0t$. MLG: impurity added on top of MLG; A: impurity located on top of sublattice A in BLG; B: impurity located on top sublattice B in BLG. } \label{Fig:basic}
\end{figure}

\begin{figure}[t]
\begin{center}
\includegraphics[scale=0.6, bb=56 200 600 380]{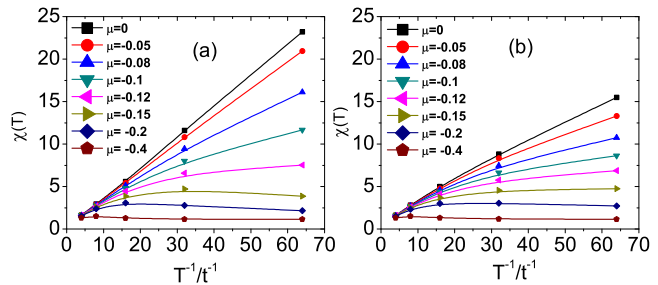}
\end{center}
\caption{(Color online). (a) The spin susceptibility $\chi$ as a function of $T^{-1}$ for case A . (b) $\chi$ as a function of $T^{-1}$ for case B . $U=1.6t$, $\varepsilon_{d}=-U/2$, $V=1.0t$.    } \label{Fig:kaiT}
\end{figure}

In Fig.~\ref{Fig:basic} we present the results of several thermodynamic quantities on impurity site as a function of $\mu$ . The error bars are smaller than the points except where shown.
In the following sections and figures, the case MLG, A and B is referred to the impurity located on top of MLG, sublattices A and B in BLG, respectively.
Since $\varepsilon_{d}=-U/2$ and $\mu=0$, the system has particle-hole symmetry. This symmetry fixes the total charge $n_d=\langle n_{d\uparrow}+n_{d\downarrow}\rangle$ exactly at one in Fig.~\ref{Fig:basic}(a). Seeing from Fig.~\ref{Fig:basic}, it is clear that the four quantities decrease when $\mu$ moves below the Dirac point. In the values of charge, there is an order: case A $>$ case MLG $>$ case B . The values of the local moment squared $m_d^2$ and the spin susceptibility $\chi$ also show the same order, while the double occupancy shows the opposite order to them. This results from the relation $2n_{\text{dup}}n_{\text{ddown}}=n_d-m_d^2$, which is satisfied automatically in our simulation although these three quantities are calculated independently. In Fig.~\ref{Fig:kaiT}, we study the spin susceptibility $\chi(T)$ versus temperature $T$ in cases A and B . In both cases, as $\mu$ moves below the Dirac point and $T$ is lowered, we see that $\chi$ crosses over from a Curie-Weiss behavior to the $T$-independent behavior as a screened local moment. In particular, near the Dirac point, we see that the local moment on the sublattice A is developed better that on the sublattice B as the temperature is lowered. In principle, spin susceptibility defined in Eq.~(\ref{eq:suscep}) depends not only on the local moment itself but also on the spin correlation with conduction-band electrons. \cite{Gubernatis87} As a result, when the LDOS of conduction electrons is higher, the spin correlation is larger, and $\chi$ has smaller value and this is consistent with the results of LDOS in Fig.~\ref{Fig:dos}. Later we will show spin correlation directly, and this point can be seen more clearly.

\begin{figure}[t]
\begin{center}
\includegraphics[scale=0.45, bb=26 20 700 420]{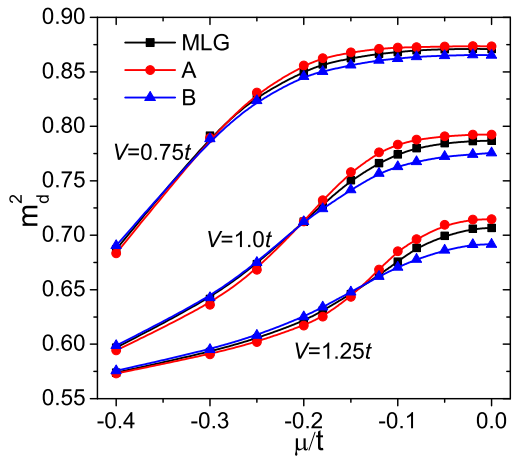}
\end{center}
\caption{(Color online). $m_{d}^{2}$ as a function of $\mu$ for different $V$, from top to down is $V=0.75t, 1.0t, 1.25t$. In all cases,
 $U=1.6t$, $\varepsilon_{d}=-U/2$, $\beta=1/T=40t^{-1}$.} \label{Fig:u16md2V}
\end{figure}

\begin{figure}[t]
\begin{center}
\includegraphics[scale=0.45, bb=26 20 600 430]{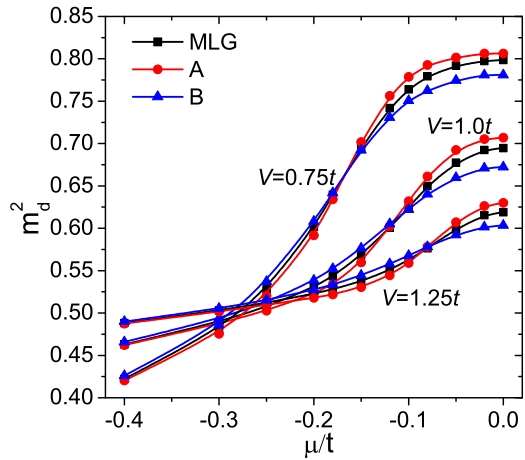}
\end{center}
\caption{(Color online). $m_{d}^{2}$ as a function of $\mu$ at $U=0.8t$ for different $V$, from top to down is $V=0.75t, 1.0t, 1.25t$. In all cases, $\varepsilon_{d}=-U/2$, $\beta=1/T=40t^{-1}$.}\label{Fig:u8md2V}
\end{figure}

\begin{figure}[t]
\begin{center}
\includegraphics[scale=0.45, bb=60 150 500 540]{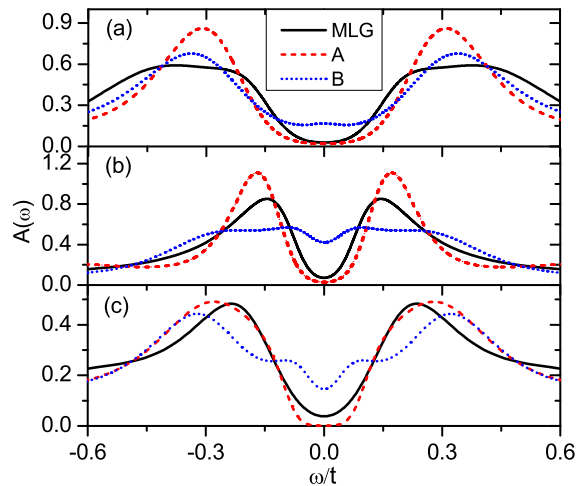}
\end{center}
\caption{(Color online). $A(\omega)$ with respect to $\omega$ for (a)$V=0.75t$, $U=1.6t$, (b)$V=1.0t$, $U=1.6t$ and (c)$V=1.0t$, $U=2.4t$. Here $\varepsilon_{d}=-U/2$, $\beta=1/T=40t^{-1}$ in all cases. } \label{Fig:spectral}
\end{figure}

In order to see how hybridization $V$ could affect the local moment, we do the calculation with different values of $V$ and the results are shown in Fig.~\ref{Fig:u16md2V}. We can see that as $V$ grows from $0.75t$ to $1.25t$, the local moment of the impurity site decreases noticeably for all the three cases we examine. As $V$ varies, the order for the three cases shown in Fig.~\ref{Fig:basic} does not change.
The Coulomb interaction $U$ for transition-metal atom in carbon-based materials can be varied from 2-5 eV, \cite{Jacob09,Wehling10,Amal10,Wehling11}
so we also study the effects of different on-site Coulomb repulsion. Shown in Fig.~\ref{Fig:u8md2V} are the results for $U=0.8t$ with other parameters be the same as those in Fig.~\ref{Fig:u16md2V}. We can see that the general behavior of the local moment remains the same with Fig.~\ref{Fig:u16md2V} as we vary $\mu$ and $V$, and the order for the cases MLG, A and B persists.
Comparing Fig.~\ref{Fig:u16md2V} with Fig.~\ref{Fig:u8md2V}, we see that the distinction for the cases A and B are more obvious with a smaller $U$.

\subsection{Spectral Densities}

In Fig.~\ref{Fig:spectral} are the spectral densities $A(\omega)$ at $1/T=40t^{-1}$ with various $U$ and $V$. We fix $\mu=0$ and $\varepsilon_{d}=-U/2$, so there is particle-hole symmetry and we have $A(-\omega)=A(\omega)$.
It is well known that for the Hartree-Fock solution of an Anderson impurity in a normal metal, \cite{Anderson61} the locations of the peaks are independent of $V$ and the separation of two peaks is about $U$. However, here we see that the separations of two side peaks in $A(\omega)$ are much smaller than Coulomb repulsion $U$, this point is consistent with the results from numerical renormalization group study. \cite{Cornaglia09}   When $V$ increases, the $A(\omega)$ peaks move toward the Dirac point. These results are in agreement with those in the previous study at higher temperature \cite{Hu11} in the case MLG , which is absolutely originated from the non-constant DOS in graphene.
We can also find that near the Dirac point, for the three cases studied, the spectral densities have the same order as that of the LDOS shown in Fig.~\ref{Fig:dos} near the Dirac point. This is because when $V$ is induced, the eigenstates in host system can greatly hybridize with impurity orbit, and as $V$ increases, the LDOS on two sublattices in graphene influences the impurity orbit more strongly.

In Fig.~\ref{Fig:spectral}(a) where $V=0.75t$ and $U=1.6t$, we can see that the differences in spectral densities in the three cases are relatively small near the Dirac point. In Fig.~\ref{Fig:spectral}(b), as $V$ is increased from $0.75t$ to $1.0t$, the peaks in $A(\omega)$ move towards the Dirac point and thus the $A(\omega)$ for case B increase dramatically while no such changes are induced in the other two cases.

We also study the $A(\omega)$ with a larger value of Coulomb repulsion $U=2.4t$ in Fig.~\ref{Fig:spectral}(c) and $V=1.0t$ with particle-hole symmetry. It is clear that the separations of two peaks in $A(\omega)$ increase but are still much smaller than $U$. Near the Dirac point, the spectral densities for the three cases have the same order as that in Fig.~\ref{Fig:spectral}(a)-(b).

\subsection{Correlation Functions}

\begin{figure}[t]
\begin{center}
\includegraphics[scale=0.6, bb=300 60 100 505]{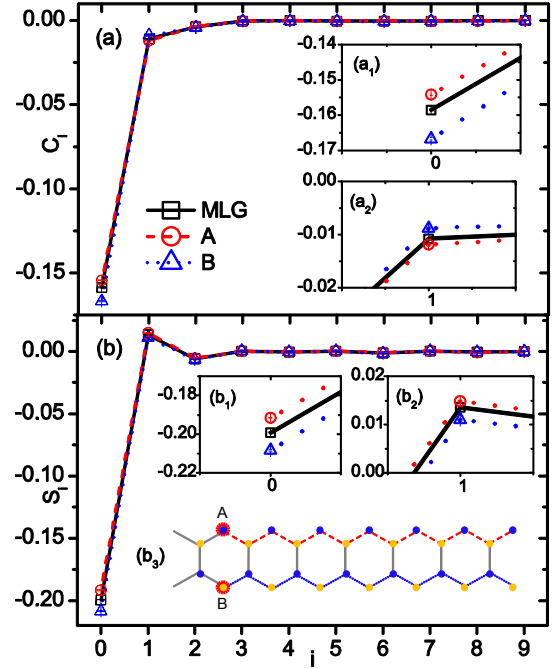}
\end{center}
\caption{(Color online). (a)The charge-charge correlation $C_i$ and (b) the spin-spin correlation $S_i$ versus site $i$. $i$ is along a zigzag direction in layer $1$ of BLG, where impurity is located on top of site $i=0$. Insets $(a_1)$ and $(a_2)$: details of $C_0$ and $C_1$; insets $(b_1)$ and $(b_2)$: details of $S_0$ and $S_1$. In inset $(b_3)$, the sites along red(dashed) and blue(dotted) zigzag lines indicate the carbon sites we consider for case A and B, respectively. The two red circled sites are the sites $i=0$, where impurity is located for cases A and B.
Here we use $V=1.0t$, $U=0.8t$ and $\varepsilon_{d}=-U/2$, $\beta=1/T=40t^{-1}$ in all the cases.} \label{Fig:corrr}
\end{figure}

In Fig.~\ref{Fig:corrr}, we present the results of charge-charge correlation $C_i$ and spin-spin correlation $S_i$ between impurity and conduction-band electrons. We set $V=1.0t$, $U=0.8t$, $\varepsilon_d=-U/2$. The chemical potential is fixed at zero point, so the system has particle-hole symmetry. In every subfigure, the impurity is located on the top of the site $i=0$, so the locations with even index are sites on the same sublattice as the site where impurity is added and those with an odd index are sites on the opposite sublattice.
As shown in Fig.~\ref{Fig:corrr}, $C_i$ and $S_i$ for all the three cases are relatively short-ranged such that the magnitudes decay rapidly with respect to the location $i$. $C_i$ lacks oscillations since at $\mu=0$ the total system is half filled and the charge exchange between two sites is greatly suppressed. If we look at the correlations in details from the insets $(a_1)$ and $(a_2)$ in Fig.~\ref{Fig:corrr}, we find that compared to the case B, the on-site correlation $|C_0|$ of the case A is weaker while the nearest-neighbor correlation $|C_1|$ is stronger.

The plots of $S_i$ show that at $\mu=0$, the impurity spin is antiferromagnetically correlated with the conduction electron spins on the same sublattice, and ferromagnetically correlated with those on the opposite sublattice.
To compare $S_i$ in the cases A and B , we also present the insets $(b_1)$ and $(b_2)$, and we see that $S_0$ is stronger for the B case while ferromagnetic correlation $S_1$ is stronger for the case A , which is consistent with $C_i$. Due to the bipartite nature of the lattice and the localization of impurity spin, the system would have weak spin fluctuation around the adatom at half filling.

\begin{figure}[t]
\begin{center}
\includegraphics[scale=0.65, bb=26 260 700 400]{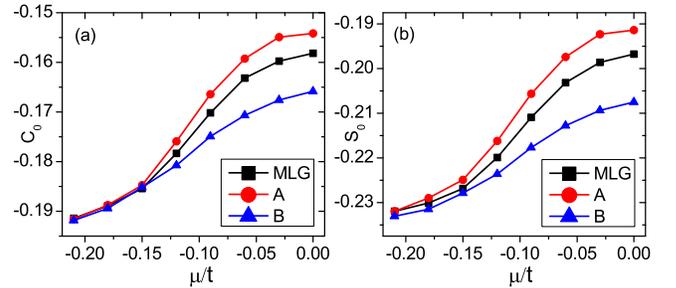}
\end{center}
\caption{(Color online). (a)The charge-charge correlation $C_0$ and (b)the spin-spin correlation $S_0$ versus $\mu$. Here we use $V=1.0t$, $U=0.8t$ and $\varepsilon_{d}=-U/2$, $\beta=1/T=40t^{-1}$.
 }
 \label{Fig:corrmu}
\end{figure}

We also focus on the behavior of on-site correlation functions $C_0$ and $S_0$ with $\mu$ moving below the zero point in Fig.~\ref{Fig:corrmu}.
The particle-hole symmetry is broken and the filling is shifted from one.  Here $V=1.0t$, $U=0.8t$ , $\varepsilon_d=-U/2$ and $\beta=40t^{-1}$.
As $\mu$ is tuned below the Dirac point, the amplitude of on-site charge correlations $C_0$ in Fig.~\ref{Fig:corrmu}(a) increase because the occupancy of the impurity orbit and conduction band shift from half filling so their charge exchange enhances.  The differences of $C_0$ for the three cases are the most obvious at $\mu=0$ and as $\mu$ is lowered, the differences become smaller and at $\mu \approx-0.2t$, three curves touch each other. We can see the same behavior of $S_0$  in Fig.~\ref{Fig:corrmu}(b) as that in $C_0$.

We can correlate the $S_i$ with the spin susceptibility in Fig.~\ref{Fig:basic}(d) and Fig.~\ref{Fig:kaiT}. At $\mu=0$ with particle-hole symmetry, all of the three cases we study have half filling, so there exists well-defined local moment on impurity site. The main differences for spin susceptibilities among them originate form  the screening of conduction-band electrons (in fact $T\chi$ is screened moment), which is reflected by the spin-spin correlations. Furthermore, $S_i$ is directly depended on the LDOS of conduction band in Fig~\ref{Fig:dos}, so in Bernal stacked BLG, the spatial inhomogeneity for spin susceptibility can be understood.

\section{Conclusions}
In summary, we have studied a magnetic impurity placed on the top of two nonequivalent sublattices in Bernal stacked BLG with Slonczewski-Weiss-McClure parameterization. The results obtained from the quantum Monte Carlo method are essentially exact, in the sense that we start with an infinite sea of conduction electrons and use no approximations to deal with many-body problem in our simulations. The LDOS on the two sublattices show spatial inhomogeneity in BLG. As a result, when we put magnetic impurity on the top of the two sublattices, such spatial inhomogeneity greatly influences the magnetic property of the impurity. It is interesting that this inhomogeneity was also seen in vacancy-induced magnetism in BLG experimentally. \cite{Ugeda10} In general, we found that the local moment on the sublattice A is conserved better than that on sublattice B and this difference becomes more apparent as hybridization $V$ increases and Coulomb repulsion $U$ decreases. Other physical quantities mostly have the same feature. The STM could be used to measure the spectral densities and the charge-charge correlation functions, and a spin-polarized STM could be used to measure the spin-spin correlations, \cite{Zhuang09,Uchoa09,Saha10} so our studies are well connected to experiments.

\section{Acknowledgement}
This work was supported by the Research Grants Council of Hong Kong (402310, HKUST3/CRF/09). F. M. Hu was supported by Academy of Finland through its Center of Excellence (2012-2017) program. We acknowledge the CPU time from CUHK in Hong Kong and CSC-IT Center for Science Ltd in Finland.

\end{document}